# Numerical Investigations of Low-Latitude Ground Magnetic Fields due to Cavity Mode Oscillations

K. Majmudar[1], R. L. Lysak[1], K. Takahashi[2]

[1]School of Physics and Astronomy, University of Minnesota, Minneapolis, MN; [2]The Johns Hopkins University Applied Physics Laboratory, Laurel, MD

Corresponding author: Khilav Majmudar (majmu008@umn.edu)

**Key points:**

- We have developed a global MHD wave code in spherical geometry that can investigate low-latitude physics.
- An MHD model with a dense plasmasphere and a height-resolved ionosphere shows cavity mode oscillations following an interplanetary shock.
- A theoretical justification is given for fast travel times of equatorial compressional waves from midnight (~2 Re) to the ground at noon.

## Abstract

Pi2 oscillations have been known to be associated with cavity mode oscillations (CMOs) in the inner magnetosphere at low latitudes. CMOs can also be excited by interplanetary shocks on the dayside. We present results from a global linear MHD wave model in spherical coordinates that allows us to analyse these waves at equatorial latitudes. We see that the model reproduces some of the principal CMO harmonics. The standing wave cavity modes are set up due to a dense plasmasphere. The CMOs associated with interplanetary shocks are compared with ground observations from the EMMA array at mid-latitudes. In accordance with observations of Pi2 pulsations, we also see a small phase difference between compressional mode Alfvén waves at midnight at low altitudes and on the noon ground, both at equatorial latitudes. This supports the finding that upon being excited in the nightside inner magnetosphere, these waves travel quickly around to the dayside. We compare our results to observations of the night-to-day time delay of ground Pi2 pulsations.

## Plain language summary

The plasma of electrons and ions and the magnetic field that surrounds the Earth reacts to disturbances originating from the Sun. These disturbances cause waves to travel along and across magnetic field lines. Some of these waves oscillate at low frequencies, which have been detected at ground stations. We aim to understand the physical basis behind the generation of these waves. To that end, we use computational methods to solve low-temperature plasma equations. These equations are set in spherical geometry, which allows us to analyze waves on the ground at equatorial latitudes. We show that waves caused by solar wind impulses on the dayside oscillate at frequencies observed by ground stations. We also show that nightside excitations cause oscillations that travel quickly from the nightside to the dayside near the Earth at equatorial latitudes, which conforms with satellite and magnetometer observations.

## 1. Introduction

Pi2 waves are low-frequency irregular oscillations with period 40-150 s. They have been observed in connection with substorms and are understood to be generated due to sudden processes in the magnetotail like bursty bulk flows due to reconnection (Angelopoulos et al., 1994) and dipolarizing flux bundles (Liu et al., 2017). Numerical modelling of these waves has usually been done using linearized MHD equations. Many previous models (Chen and Hasegawa, 1974; Lee and Lysak, 1991; Streltsov and Lotko, 1999) have followed the geometry of the Earth's dipole magnetic field, with one coordinate along the background field and the other two being orthogonal to it. Such a geometry is a natural setting for analysis for equations in the Earth's inner magnetosphere. However, adjustments must be made to incorporate the spherical geometry of the ionosphere if one is interested in including its effects. One such adjustment introduces nonorthogonality in the dipole grid, as in Lysak (2004) and Lysak et al. (2013). In this study, we instead model the wave equations using a spherical coordinate system.

Dipole grids contain a singularity at the equator due to vanishingly short field lines. Thus, models based on a dipole grid can be used at best till midlatitudes. This introduces an additional complication of a truncation of the spherical harmonic expansion below the ionosphere, as the grid does not cover an annulus around the equator (Appendix A in Lysak, 2004). A capability of spherical models is that they can model low latitude ($< 30°$) and equatorial geomagnetic fields. The presence of ground-based magnetometer stations near equatorial latitudes makes it possible to compare results from such numerical models with observational data, such as is presented in Takahashi et al. (2024). There remains the issue of the singularity at the poles, which is dealt with in a manner shown in section 2.

We have also used this model to analyze the radial structure of compressional waves, termed cavity mode oscillations (CMOs). CMOs are amplified at eigenmodes of the fast mode in the inner magnetosphere. They may be excited by the same processes that are believed to be behind the generation of Pi2 pulsations, for example bursty bulk flows. We may imagine that at low latitudes, sudden irregular processes in the magnetotail may produce resonant CMOs at ULF frequencies in the inner magnetosphere, which in turn give rise to Pi2 waves observed on the ground at low latitudes. CMOs can also be produced by interplanetary shocks. Sudden impulses due to strong solar events like coronal mass ejections excite CMOs that have been studied through multiple spacecraft observations (Takahashi et al., 2018). Shock-induced CMOs are easier to distinguish from other ULF waves that are continuously present (Hartinger et al., 2013). Therefore, structures in the inner magnetosphere such as CMOs (also called plasmaspheric virtual resonances by Lee, 1998 and Lee and Lysak, 1999) are believed to play an important role for Pi2 generation at low latitudes. The possibility of Pi2 pulsations being caused by BBFs via compressional magnetic pulses in the magnetosphere formed due to the braking of these flows was proposed by Shiokawa et al. (1998). Other studies (Kepko et al., 2001; Kepko and Kivelson, 1999) lent support to the possibility of BBFs being the source mechanism for Pi2 generation. At high to mid latitudes, BBFs may influence the generation of Pi2 pulsations on the ground more directly (Panov et al., 2010, 2013, 2014).

In this paper we first introduce the model in section 2 with the relevant MHD equations cast in spherical geometry with details of the density structure used. In section 3 we analyse the time history and spectral structure of waves at Pi2 frequencies generated by the impact of an interplanetary shock at various points in the dayside magnetosphere and the ground and compare our results to observational data presented in Takahashi et al. (2018). In section 4 we will look at compressional mode Alfven waves at low latitudes driven by bursty bulk flows on the nightside.

## 2. Model Description

We use the following equations of linear MHD:

$$\begin{gathered}\partial_t \mathbf{b} = -\nabla \times \mathbf{E} \\ \partial_t \mathbf{v} = \frac{1}{\mu_0 \rho} (\nabla \times \mathbf{b}) \times \mathbf{B}_0 \\ \mathbf{E} = -\mathbf{v} \times \mathbf{B}_0\end{gathered} \tag{1}$$

We will use a right-handed coordinate system where $\hat{\mathbf{x}} = \hat{\mathbf{r}}$, $\hat{\mathbf{y}} = \hat{\varphi}$, and $\hat{\mathbf{z}} = -\hat{\theta}$ (note that $\hat{\mathbf{z}}$ points toward the north). The $\hat{z}$ and $\hat{y}$ directions are referred to by the Greek letters $\hat{\lambda}$ and $\hat{\phi}$ respectively in the plots in this paper. Here the background magnetic field can be written as $\mathbf{B}_0 = \mathrm{B}_0(\hat{\mathbf{x}}\cos\chi + \hat{\mathbf{z}}\sin\chi)$, where $\mathrm{B}_0 = (\mathrm{B}_{\mathrm{eq}}/\mathrm{r}^3)\sqrt{1+3\cos^2\theta}$, $\cos\chi = -2\cos\theta/\sqrt{1+3\cos^2\theta}$ and $\sin\chi = \sin\theta/\sqrt{1+3\cos^2\theta}$. Then we have

$$
\begin{aligned}
E_x &= -v_y B_0 \sin\chi \\
E_y &= (v_x \sin\chi - v_z \cos\chi) B_0 \\
E_z &= v_y B_0 \cos\chi
\end{aligned}
\tag{2}
$$

The equations of motion then become:

$$
\begin{aligned}
\partial_t v_x &= \frac{B_0 \sin\chi}{\mu_0 \rho} (\nabla \times \boldsymbol{b})_y \\
\partial_t v_z &= -\frac{B_0 \cos\chi}{\mu_0 \rho} (\nabla \times \boldsymbol{b})_x \\
\partial_t v_y &= \frac{B_0}{\mu_0 \rho} [(\nabla \times \boldsymbol{b})_z \cos\chi - (\nabla \times \boldsymbol{b})_x \sin\chi]
\end{aligned}
\tag{3}
$$

Writing these in terms of the electric field

$$
\begin{aligned}
\partial_t E_x &= V_A^2 \sin\chi\, [(\nabla \times \boldsymbol{b})_x \sin\chi - (\nabla \times \boldsymbol{b})_z \cos\chi] \\
\partial_t E_y &= V_A^2 (\nabla \times \boldsymbol{b})_y \\
\partial_t E_z &= V_A^2 \cos\chi\, [(\nabla \times \boldsymbol{b})_z \cos\chi - (\nabla \times \boldsymbol{b})_x \sin\chi]
\end{aligned}
\tag{4}
$$

Now we rotate these equations into a field-aligned system by writing

$$
\begin{aligned}
E_\perp &= E_x \sin\chi - \mathrm{E}_z \cos\chi \\
E_\parallel &= E_x \cos\chi + E_z \sin\chi = 0
\end{aligned}
\tag{5}
$$

Parallel electric fields are neglected owing to the large scale size of the system. This scale size also allows us to ignore kinetic and electron inertial effects. Note that $E_\perp$ points inward, i.e., in the $\hat{\boldsymbol{b}} \times \hat{\varphi}$ direction. So we can write

$$
\partial_t E_\perp = V_A^2 [(\nabla \times \boldsymbol{b})_x \sin\chi - (\nabla \times \boldsymbol{b})_z \cos\chi] = V_A^2 (\nabla \times \boldsymbol{b})_\perp \tag{6}
$$

Now we can add the Ohmic currents in the same manner. We write, again neglecting parallel electric fields:

$$
\boldsymbol{j}_I = \sigma_P (\widehat{\boldsymbol{B}}_0 \times \boldsymbol{E}) \times \widehat{\boldsymbol{B}}_0 + \sigma_H (\widehat{\boldsymbol{B}}_0 \times \boldsymbol{E}) \tag{7}
$$

Note here that $\widehat{\boldsymbol{B}}_0 = \hat{\boldsymbol{r}} \cos\chi + \hat{\theta} \sin\chi$, we can write the ionospheric current components

$$
\begin{aligned}
j_{Ix} &= \sigma_P \sin\chi\, (E_x \sin\chi - E_z \cos\chi) - \sigma_H E_y \sin\chi \\
j_{Iy} &= \sigma_P E_y + \sigma_H (E_x \sin\chi - E_z \cos\chi) \\
j_{Iz} &= -\sigma_P \cos\chi\, (E_x \sin\chi - E_z \cos\chi) + \sigma_H E_y \cos\chi
\end{aligned}
\tag{8}
$$

As in the case with the electric fields in equation 5 above, we can define $j_{I\perp} = j_{Ix} \sin\chi - j_{Iz} \cos\chi$, so we have

$$j_{I\perp} = \sigma_P E_\perp - \sigma_H E_y$$

$$j_{Iy} = \sigma_P E_y + \sigma_H E_\perp \tag{9}$$

Subtracting these from the evolution equation for the electric field, we get

$$\partial_t E_y = V_A^2 (\nabla \times \mathbf{b})_y - \frac{\sigma_P}{\varepsilon_\perp} E_y - \frac{\sigma_H}{\varepsilon_\perp} E_\perp$$

$$\partial_t E_\perp = V_A^2 (\nabla \times \mathbf{b})_\perp - \frac{\sigma_P}{\varepsilon_\perp} E_\perp + \frac{\sigma_H}{\varepsilon_\perp} E_y \tag{10}$$

The basic scheme is to calculate the curl of **E** in spherical coordinates to advance **b** as in equation (1), then determine the curl of **b** in spherical coordinates, and rotate it to the field-aligned coordinates in the same manner as equation (5), and use it in equation (10) to advance the perpendicular components of the electric field. Then using the inverse of equation (2), and setting the parallel electric field to zero, we have $E_x = E_\perp \sin\chi$ and $E_z = -E_\perp \cos\chi$.

The fields and currents are placed in a staggered Yee grid. The singularity at the poles in a spherical coordinate system is dealt with using a prescription given by Soriano et al. (2005). The time evolution of $b_x$, which is defined at even points on the $\theta$ grid, is determined by integrating $E_y$ according to Faraday's law at the second-to-last points (which are odd) near both poles. Only $E_y$ enters the picture as $E_z$ is defined on even $\theta$ values.

$$\frac{d}{dt}\left[r^2 \int d\varphi dz b_x\right] = \pm \oint r \sin\theta \, d\varphi E_y \tag{11}$$

We include a height-resolved ionosphere in our model. Pedersen and Hall conductivities are calculated by the following formulae (Kelley, 2009; Lysak et al., 2013):

$$\sigma_P = \sum_s \frac{n_s q_s^2}{m_s} \frac{\nu_s}{\nu_s^2 + \Omega_s^2}$$

$$\sigma_H = \sum_s \frac{n_s q_s^2}{m_s} \frac{\Omega_s}{\nu_s^2 + \Omega_s^2} \tag{12}$$

Here, $\nu_s$ and $\Omega_s$ respectively refer to the collision frequency and gyrofrequency of species $s$. The ion composition is obtained from the same reference (Kelley, 2009). Densities are calculated using the Chapman layer formula for a dayside ionosphere (Schunk and Nagy, 2000):

$$n(z) = n_0 \exp\left[1 - \frac{z - z_0}{h} - e^{\frac{z - z_0}{h}}\right] \tag{13}$$

The scale height $h$ below the peak height $z_o$ is allowed to be different from that above the peak. The ion composition is modulated according to Table B.2 of Kelley (2009). Density and Alfven speed profiles are shown in Figure 1.

The lower boundary of the model is fixed at 250 km for some runs (figures 2-6) and at 100 km for others (figures 7-9). This does not make a noticeable difference in most cases. In the 100 km case the density structure of the ionosphere leads to large gradients in the Alfven speed which can cause numerical instabilities. This issue gets mitigated with a lower boundary at 250 km but this also lowers the peak value of the Alfven speed. The residual conductivities below the bottom height are packaged into height-integrated Pedersen and Hall conductances. These height-integrated quantities are used in the jump conditions at this lower conducting sheet boundary, which couple the magnetic field in the insulating atmosphere to the rest of the simulation domain:

$$\hat{r} \cdot \Delta b = 0$$

$$\hat{r} \times \Delta b = \mu_0 \underline{\Sigma} \cdot E \tag{14}$$

Here, $\underline{\Sigma}$ represents the height-integrated conductance tensor (Lysak, 2004; Sciffer and Waters, 2002). Below the ionosphere boundary, the atmosphere is assumed to be electrically insulating. Consequently, the magnetic field here is computed as the gradient of a magnetic scalar potential (Lysak, 2004). The scalar potential satisfies Laplace's equation, $\nabla^2 \psi = 0$. Decomposing $\psi$ in a spherical harmonic expansion in spherical coordinates, we can calculate ground magnetic fields using $b = \nabla\psi$. We assume a perfectly conducting ground, which implies that the radial component of $B$ vanishes there. This gives us the boundary condition $b_x(r_E) = 0$. Recall that in our coordinate system, $\hat{x}$ denotes the radial component, which on the ground is the vertical component. This is often denoted by $Z$, completing the ground coordinate system with the $H$ (north-south) and $D$ (east-west) components.

The outer boundary of the simulation is at 8 Re. Here the simulation is initiated by an electric field in the form of a downgoing wave. The incoming nature of the wave is set by adding a term with the ratio of the electric-to-magnetic fields set to the local Alfven speed:

$$E_y = E_y^{drive} + V_A b_z \tag{15}$$

The grid has 64 cells in the longitudinal ($\hat{y}$) direction and 256 in the latitudinal ($\hat{z}$) direction. In the radial ($\hat{x}$) direction, grid size and spacing is decided by the requirement that we would like to have smaller grid spacing in the regions of high Alfven speed gradients, i.e., near the ionosphere, and larger spacing where the Alfven speed gradient is weaker. This is achieved by the following function, with $r_{trans}$ signifying the radial distance of the grid-size transition:

$$\delta r = 0.5(r_{min} + r_{max}) + 0.5(r_{max} - r_{min}) tanh\left(\frac{r - r_{trans}}{\Delta r}\right) \tag{16}$$

As an example, for an outer boundary at 8 Re with $r_{trans} = 2.5$ and $\Delta r = 0.1$, we get 339 cells.

## 3. Cavity mode oscillations and their ground signatures

We compare values of ground magnetic fields obtained from our model with data of the $B_\lambda$ or horizontal H (northward) and $B_\phi$ or D (eastward) components from geomagnetic observations by the European quasi-Meridional Magnetometer Array (EMMA). The EMMA observations were taken following an interplanetary shock in the dayside magnetosphere that excited cavity mode oscillations. We are referring here to figure 8 of the paper (Takahashi et al., 2018). At the outer boundary, we use a driving electric field with a Gaussian spatial form:

$$\tilde{E} = E_0 e^{-\alpha(\lambda-\lambda_0)^2} e^{-\beta(\phi-\phi_0)^2} \tag{17}$$

Here, $\lambda$ and $\phi$ are respectively the usual latitudinal and longitudinal angles and $\alpha$ and $\beta$ are positively valued scale factors. $\lambda_0$ and $\phi_0$ specify the locations of the peaks of the profiles. This driver is applied on the dayside boundary. We use a few different types of time driving profiles. The first is a linear ramp to simulate a shock.

$$E_{drive} = \tilde{E}\frac{t}{t_1} \tag{18}$$

This profile is applied up until an amount of time $t_1$, after which $E_{drive} = \tilde{E}$ for the rest of the simulation. For the results shown in this paper, $t_1 = 50\ s$ and $E_0 = 20\ mV/m$. Angelopoulos et al. (1994) mention that the time scale of flow bursts if of the order of one minute, which justifies our choice for the ramp length. Consider an order-of-magnitude calculation where $E_\phi/B_\lambda \sim V_{xF}$ , where $V_{xF}$ is the radial flow velocity of the bursty flows. This velocity has been shown to be about 500 km/s to 1000 km/s in Angelopoulos et al. (1994). Coupled with a magnetic field intensity of ~30 nT, this gives us an electric field in a similar order of magnitude as our $E_0$. Similar conclusions arise from THEMIS data presented by Panov et al. (2010), which gives an $E_0$ larger than the one in our study but within the same order of magnitude. The second type of driving profile used is a collection of sine waves up to 50 mHz with random phases. This driver allows us to see which frequencies are the resonant frequencies that the system picks out. We refer to this as the random wave driver in our figure captions. It simulates the oscillatory impact of a bursty bulk flow.

It should be noted that the respective driving fields for all the plots mentioned in this section, i.e., Figures 2-6, were located at the equatorial noon boundary with $\lambda_0 = 0°$ and $\phi_0 = 180°$. The widths of the Gaussians are given by $\sqrt{\alpha}$ and $\sqrt{\beta}$, with $\alpha = (1/15°)^2$ and $\beta = (1/45°)^2$. Ground signatures from our runs are shown in figure 3 and the top panel of figure 2. The oscillations we are interested in are those of period 30 s – 40 s, or 25 mHz – 34 mHz. We see that the horizontal northward (H) component of the ground magnetic field (figure 3) is generally of a larger magnitude than the eastward (D) component (top panel of figure 2). In the study of Takahashi et al. (2018), the ground magnetometers are situated roughly between MLT 10 and 12. Therefore we choose to plot our ground signatures at MLT 11. Our latitudes are varied according to the L values of the ground stations in Figure 8 of Takahashi et al. (2018). We see that the time series gains more structure with increasing latitude. A characteristic ring in the beginning, corresponding to the immediate effects of the ramp, is also seen. This is a marker of the "irregularity" in Pi2 oscillations, which arises due to fast and sudden impulses. The bottom panel of Figure 2 shows time series of the compressional $B_\lambda$ at equatorial footpoints of two of the ground fields. They increase in magnitude throughout the length of the run as energy is introduced into the system by the ramp boundary condition, which settles to a constant value after time $t_1$.

Resonant frequencies of the cavity mode oscillations at $r = 2.5$ are shown in Figure 4. The top panel results from a random wave driver and the bottom one from the ramp driver. We see that harmonics around 25 mHz and 35 mHz are supported by the system both in the magnetosphere and on the ground. Some CMOs following a dayside interplanetary shock have been observed to oscillate at a fundamental frequency of 13 mHz and a second harmonic at 26 mHz (Takahashi et al., 2018). A correspondence between the spectral peaks of the azimuthal electric field component ($E_\phi$) at the RBSP-B location and those of the geomagnetic field was observed. Therefore, Figure 4 contains plots of the spectrum of $E_\phi$ close to the RBSP-B location in that study. It was also mentioned that different stations observed different spectral peaks of the geomagnetic field. We notice a similar correspondence, with peaks predominantly for 25 mHz and 35 mHz with the sporadic presence of the 10-15 mHz and 40 mHz modes. These peaks are seen in plots of the spectra of the ground magnetic field components in response to the random wave driver, which are shown in Figures 5 and 6. All spectra in figures 4, 5 and 6 are calculated as the magnitude of the Discrete Fourier Transform (DFT) of the field time series which is normalized by the number of samples. This retains the original field units (nT). The plotted quantity is therefore the magnitude spectrum. The four latitudes chosen in these figures are those corresponding to some of the stations in Takahashi et al. (2018), which shows data from EMMA magnetometers in the mid-latitudes.

## 4. Pi2 waves and their ground signatures

Low-latitude Pi2 pulsations have long been thought as being the result of cavity mode oscillations in the inner magnetosphere, which in turn arise in response to sudden magnetotail events like bursty bulk flows. Modelling studies have suggested a dense plasmasphere as the reason for wave trapping which sets up the CMO resonances (Allan et al., 1996; Lee, 1998; Lee and Lysak, 1999). Observational studies of mid and low latitude Pi2 waves support this hypothesis (Takahashi et al., 1995; Yeoman and Orr, 1989; Takahashi et al., 2024). Results of this section, shown in Figures 8-10, were obtained from simulation runs driven from the midnight equatorial boundary with $\lambda_0 = \phi_0 = 0°$. α, β and $t_1$ are the same as in the previous section for the run shown in Figure 8.

Compressional fast mode waves have been found to travel quickly from nightside inner magnetospheric regions at equatorial latitudes to the ground around noon at the equator (Takahashi et al., 2024). This was shown by a small phase difference between the compressional magnetic field ($B_\lambda$) measured by RBSP within 5º of the equator with L-shell values between 2 and 3.5, and the northward H component near the dayside equator measured by ground magnetometer stations (Figures 11-c and 11-h in the same reference). We have reproduced the plots in question as Figure 7 in this paper. We show our simulation result for the same metric in Figure 8, which was driven using the ramp driver. It is calculated by subtracting the phase at noon from the phase at midnight. We see that apart from the spike near zero, our phase difference is within the $\pm 30°$ limit that was observed. In the Pi2 frequency range of 6.7-40 mHz, we see a peak at 23.3 mHz. At this frequency we have a phase difference of -26.6º between the compressional wave travelling from the midnight to the noon sector. This phase can be estimated using frequency-band averaging following Bendat and Piersol (1971) and Green (1976). Averaging over 7 frequency bins (effective resolution ≈ 11.7 mHz), the coherence at 23.3 mHz is 0.94 and the phase difference is $-11°$ with a 95% confidence half-width of $11°$ (the estimate remains between $-11°$ and $-12°$ for band widths of 5 to 9 bins). The neighbouring frequencies of 21.6 mHz and 25 mHz have phase differences of -3.5º and -7.1º respectively. This peak at 23.3 mHz can be seen in the middle and bottom panels of the same figure, which are plots of the base-10 logarithms of the magnitude spectra (calculated as the magnitude of the DFT) of the compressional $B_\lambda$ at midnight and the ground $B_{\lambda}$ at noon, respectively. This mode corresponds to propagation from nightside to dayside.

Figure 9 shows plots of a radial slice of the mode structure of the azimuthal electric field ($E_{\phi}$) and the compressional magnetic field. The driver for the runs resulting in Figures 9 and 10 is a set of waves up to 50 mHz with random phases. We see waves in the Pi2 frequency range are excited with the clearest peak at 25 mHz. A node for ($B_{\lambda}$) is seen at around 3.75 Re. This point is just inside the plasmapause, which supports the presence of cavity mode oscillations as standing waves in the inner magnetosphere resulting from the presence of a dense plasmasphere. An azimuthal slice of ($B_{\lambda}$) is shown in Figure 10. We see that the 25 mHz mode is present at all MLTs and is the one with the most power.

The low phase difference suggests that the cavity mode oscillations may be oscillating coherently with the ionospheric currents included in our model. This would make the travel time of compressional wave pulses near instantaneous, resulting in the small phase shift.

## 5. Discussion and conclusions

We have thus demonstrated here results from the first MHD model which couples waves in the inner magnetosphere (L < 8) to ground pulsations at all latitudes. This unique feature is achieved by casting the equations in spherical coordinates, which renders equatorial latitudes amenable to analysis. The inclusion of a height resolved ionosphere allows us to incorporate ionospheric effects, which influence the coupling of bulk waves with ground fields.

Under the influence of different kinds of driving profiles we have obtained various properties of the system. Simulating an interplanetary shock using a ramp profile has allowed us to reproduce salient features of ground pulsations, like their structure in time and frequency space.

The inclusion of ionospheric effects offers a possible explanation for the observed quick travel time of the compressional component of the magnetic field perturbation from the nightside inner magnetosphere to the dayside noon ground, both at the equator. In both this case and the above, we saw the Pi2 frequencies in play.

There are areas in which the model needs to be improved. The ionosphere and plasmasphere components of the model could be made more realistic by the inclusion of empirical particle densities, say from the IRI and MSIS models for the ionosphere and the Global Core Plasma Model of Gallagher et al. (2000) for the plasmasphere. This would give also give us more realistic representations of interhemispheric and dayside-nightside differences. There is also the issue of not having a clear peak around 13 mHz for low latitude pulsations, which has been observed by various studies (Takahashi et al. 2018; Thomas et al., 2019). We observe in repeated runs the first prominent peak at 25 mHz. We believe that the inclusion of realistic, empirically derived particle densities should fix this problem. It has been conjectured that dayside Pi2 ground pulsations are driven by ionospheric currents that couple to nightside CMOs at Pi2 frequencies (Takahashi et al., 2024; Thomas et al., 2019). This has been tested numerically by Imajo et al. (2017) using a global current model. Future work on our code could include analysis of ionospheric current dynamics to investigate this question from a magnetohydrodynamic perspective.

Another important direction that can be explored is the inclusion of a finitely conducting ground. This would change the ground boundary conditions and would allow us to calculate ground electric fields through an impedance tensor as in, for example, Boteler (2014). These geoelectric fields allow us to quantify geomagnetically induced currents (GICs). We may then be able to quantify impacts on ground systems of Pi2 waves generated by geomagnetic events.

**Acknowledgements**

This research was supported by the NSF grant AGS2225270 to the University of Minnesota – Twin Cities. Computing resources were provided by the Minnesota Supercomputing Institute. KT acknowledges the support of the NASA grant 80NSSC21K0453.

**Conflict of Interest**

The authors declare no conflicts of interests.

**Open Research Section**

Some plots for this paper were made using Matplotlib version 3.10.7 (Hunter, 2007) which is under the PSF license, and others were made using IDL® V.9.0. The simulation code was written in Fortran 90. All the routines mentioned above are available at the Data Repository for the University of Minnesota (DRUM) (Majmudar et al., 2025) under the Attribution-NonCommercial-NoDerivatives 4.0 International license.

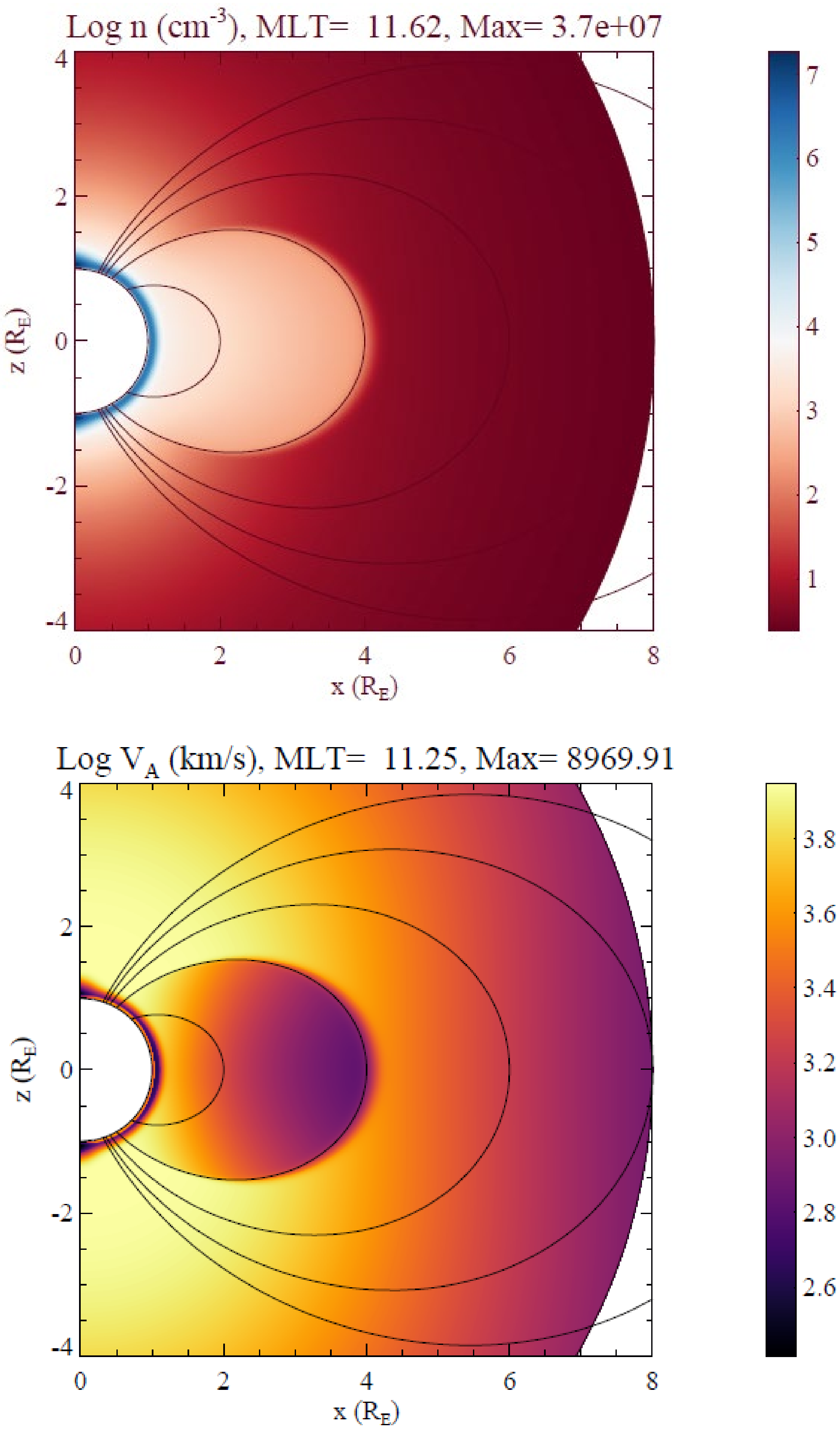

**Figure 1.** (top) Density profile for our runs. (bottom) Slice of Alfvén speed profile at midnight.

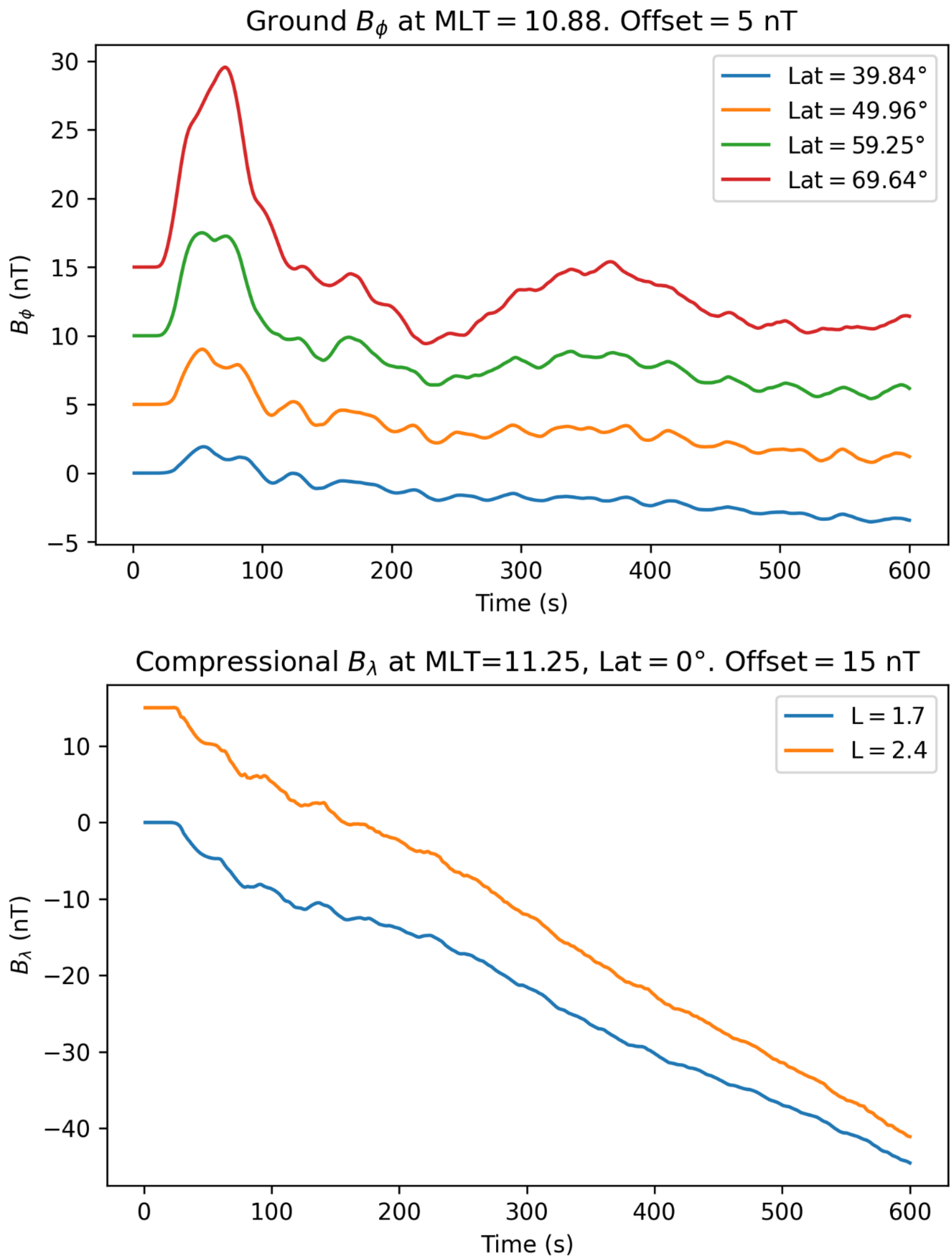

**Figure 2.** (Top) Ground $B_\phi$ (D, eastward) time series. (Bottom) Compressional $B_\lambda$ at the equatorial footpoints at the same L-shell as the ground $B_\phi$ at latitudes 39.84° and 49.96°. Driver for both plots is the ramp.

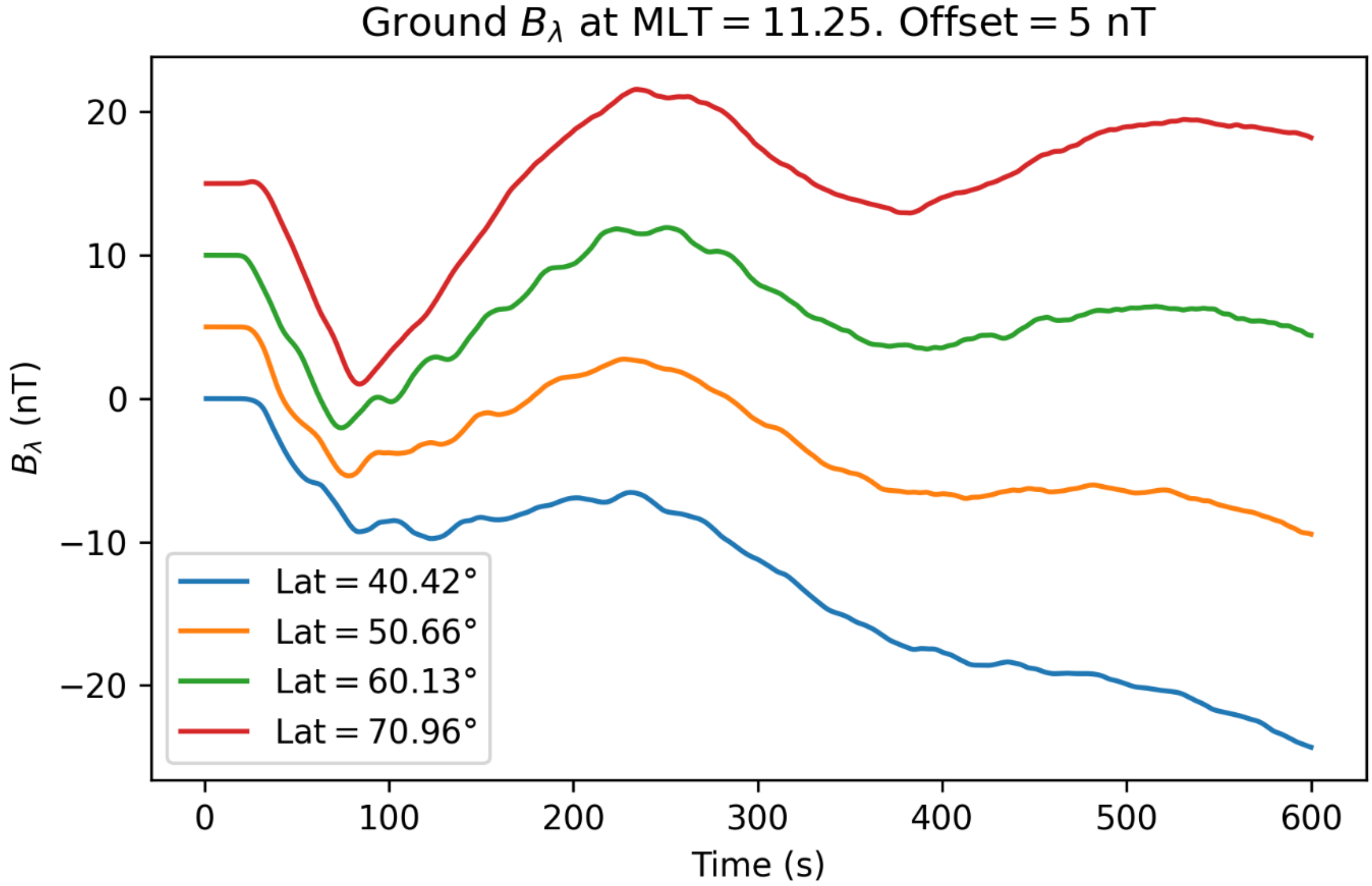

**Figure 3.** Ground $B_\lambda$ (H, northward) time series. Driver is the ramp.

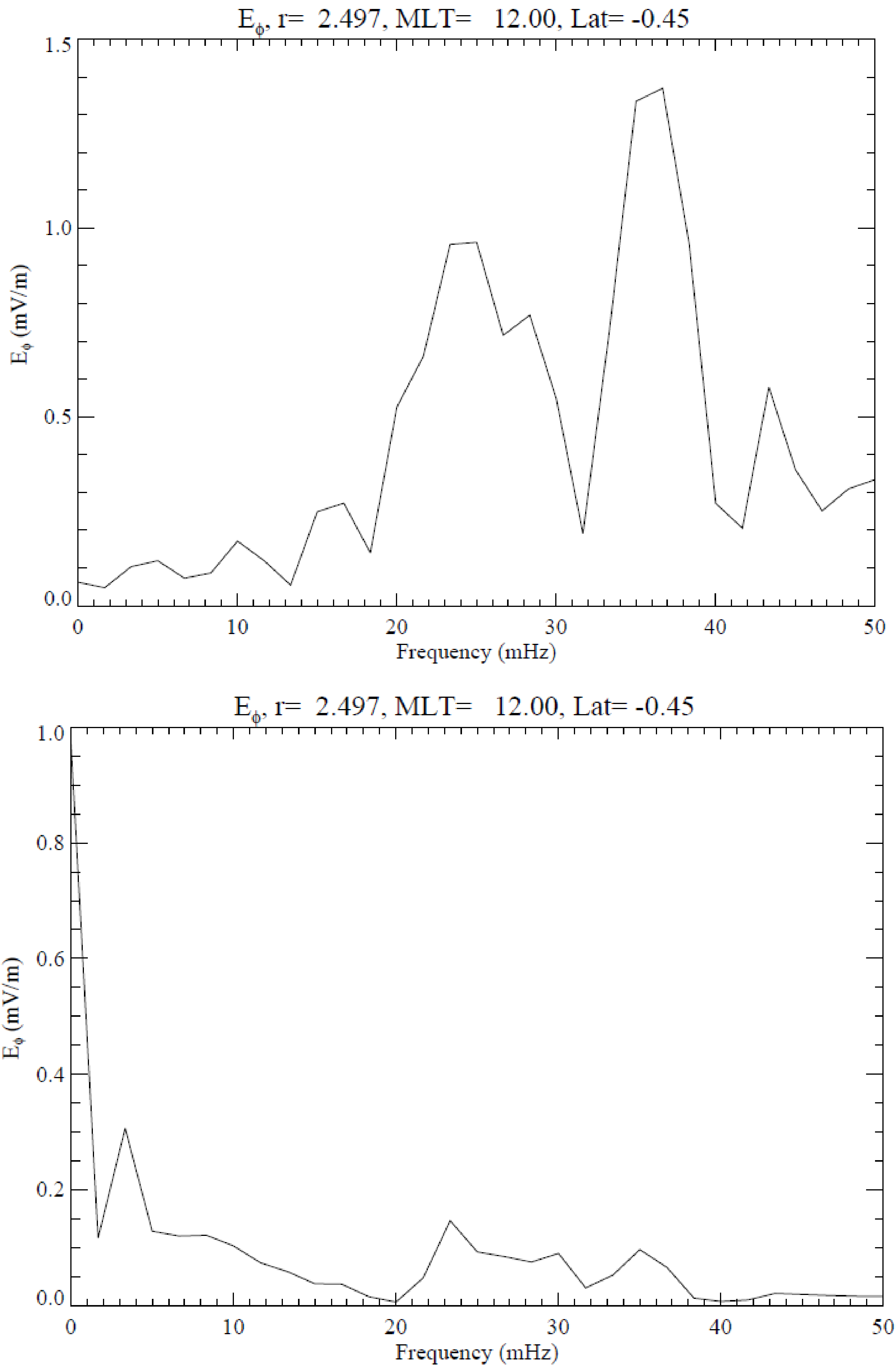

**Figure 4.** Magnitude spectrum of the poloidal $E_\phi$ component. (Top) Random wave driver (Bottom) Ramp driver.

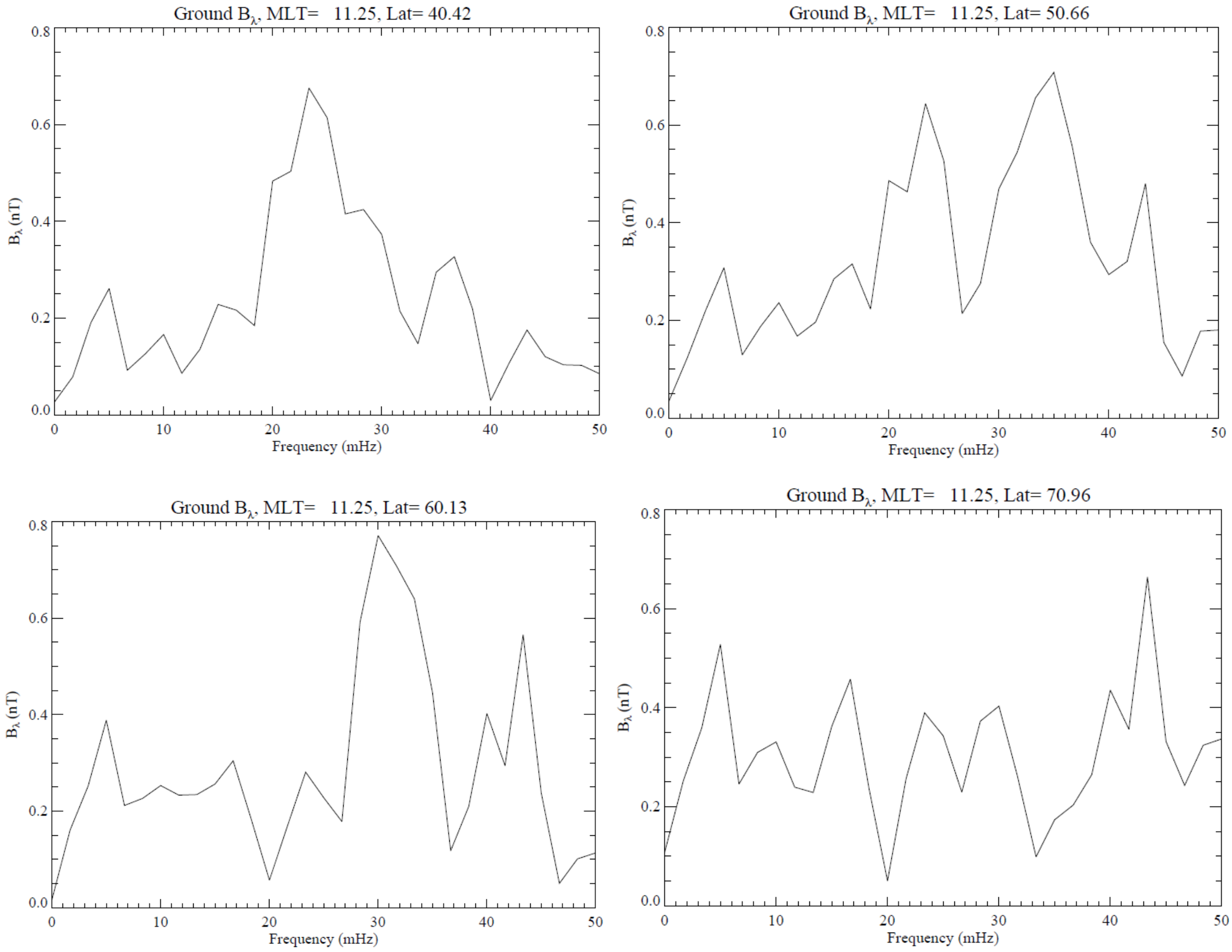

**Figure 5.** Magnitude spectrum of the ground $B_\lambda$ (H, northward) component of the geomagnetic field.

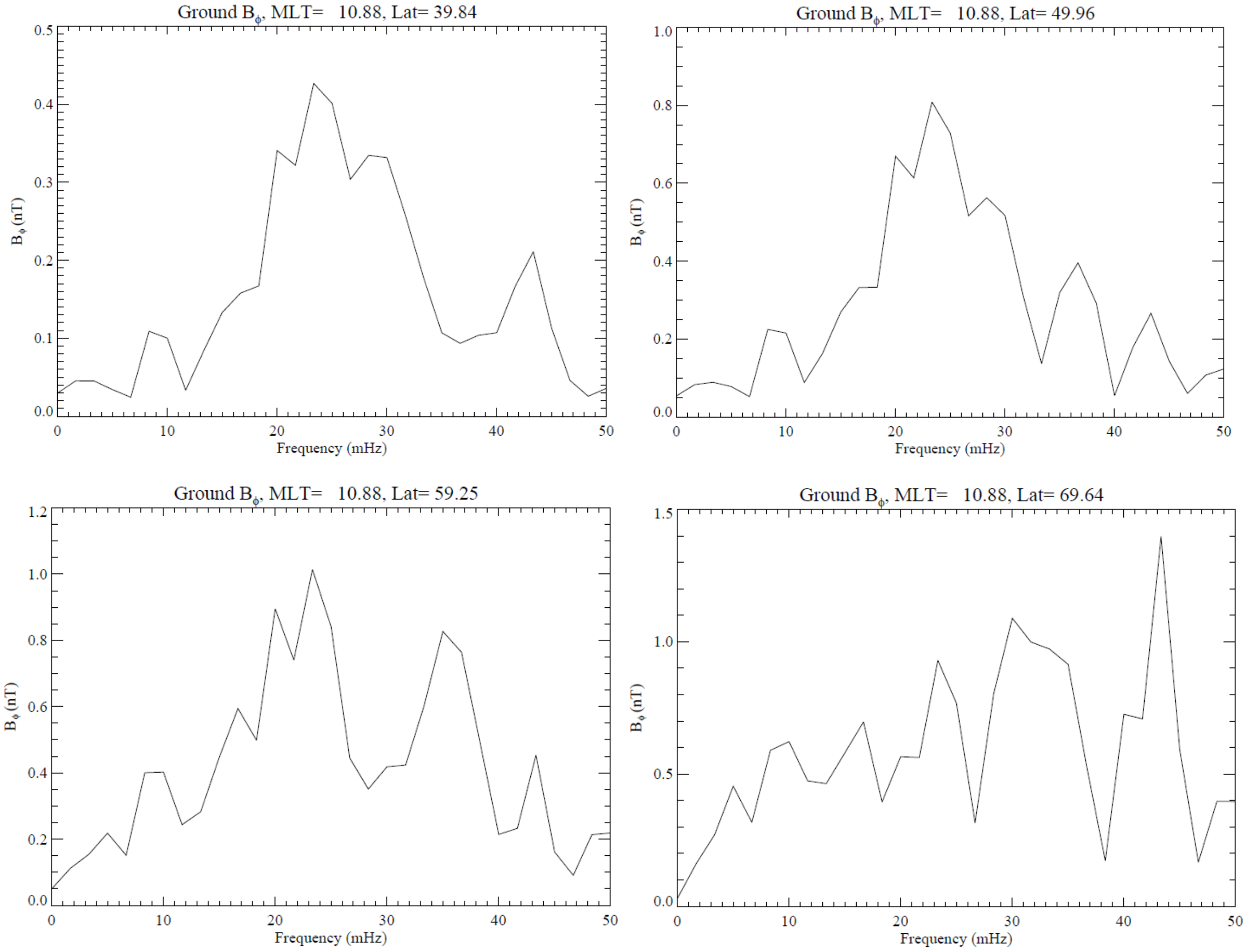

**Figure 6.** Magnitude spectrum of the ground $B_\phi$ (D, eastward) component of the geomagnetic field.

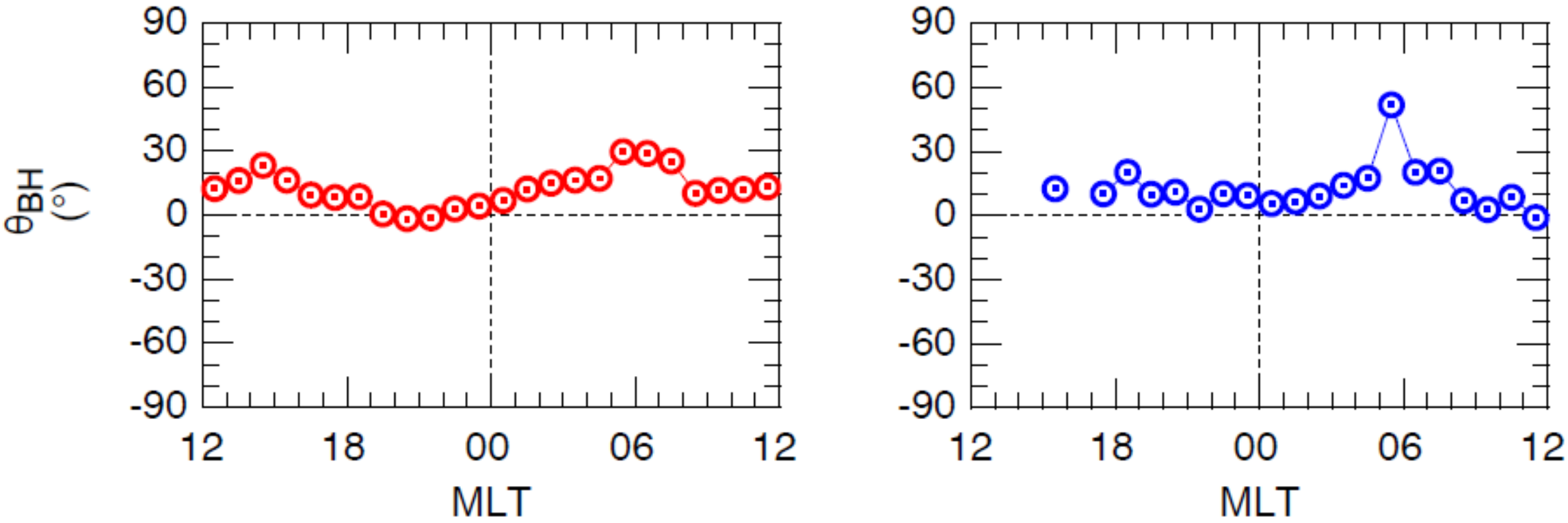

**Figure 7.** A reproduction of Figures 11-c (left) and 11-h (right) from Takahashi et al. (2024). They show the median values of MLT dependence of the phase difference of ground pulsations associated with strong (left) and weak (right) events ($B_\lambda$) detected by RBSP at $L$ = 2–3.5, within 5° of the dipole equator and within 2 hr of midnight.

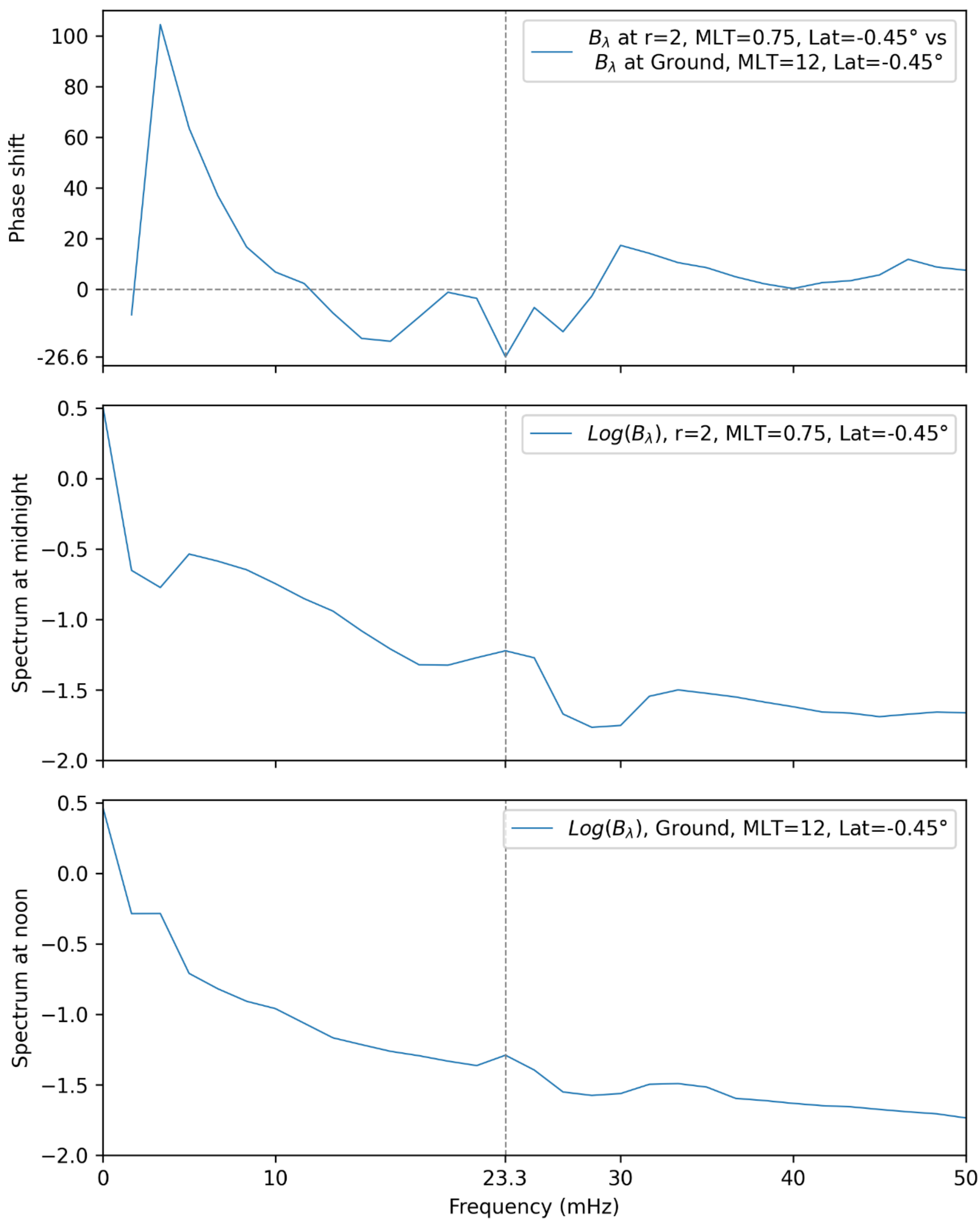

**Figure 8.** (Top) Phase difference calculated as phase(midnight) – phase(noon). (Middle) $Log_{10}$ of the magnitude spectrum of the compressional component in the magnetosphere. (Bottom) $Log_{10}$ of the magnitude spectrum of the ground $B_z$ component. Our peak of interest is 23.3 mHz, at which the phase difference is -26.6º (band averaged $-11° \pm 11°$ at 95% confidence). Driver is the ramp.

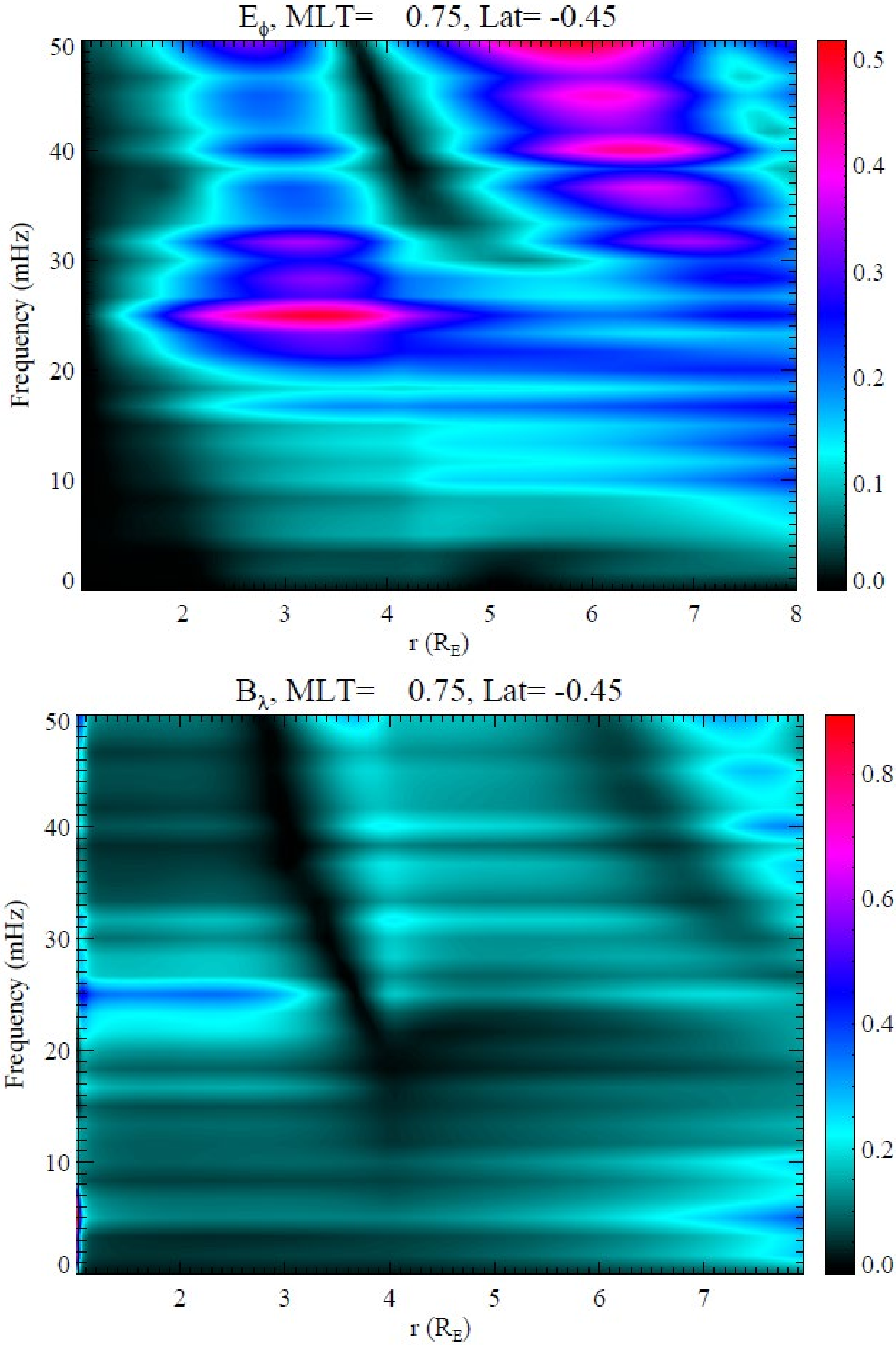

**Figure 9.** Radial slice of the mode structure of the (top) $E_\phi$, and (bottom) compressional $B_\lambda$ components. Driver for both is a collection of waves up to 50 mHz with random phases.

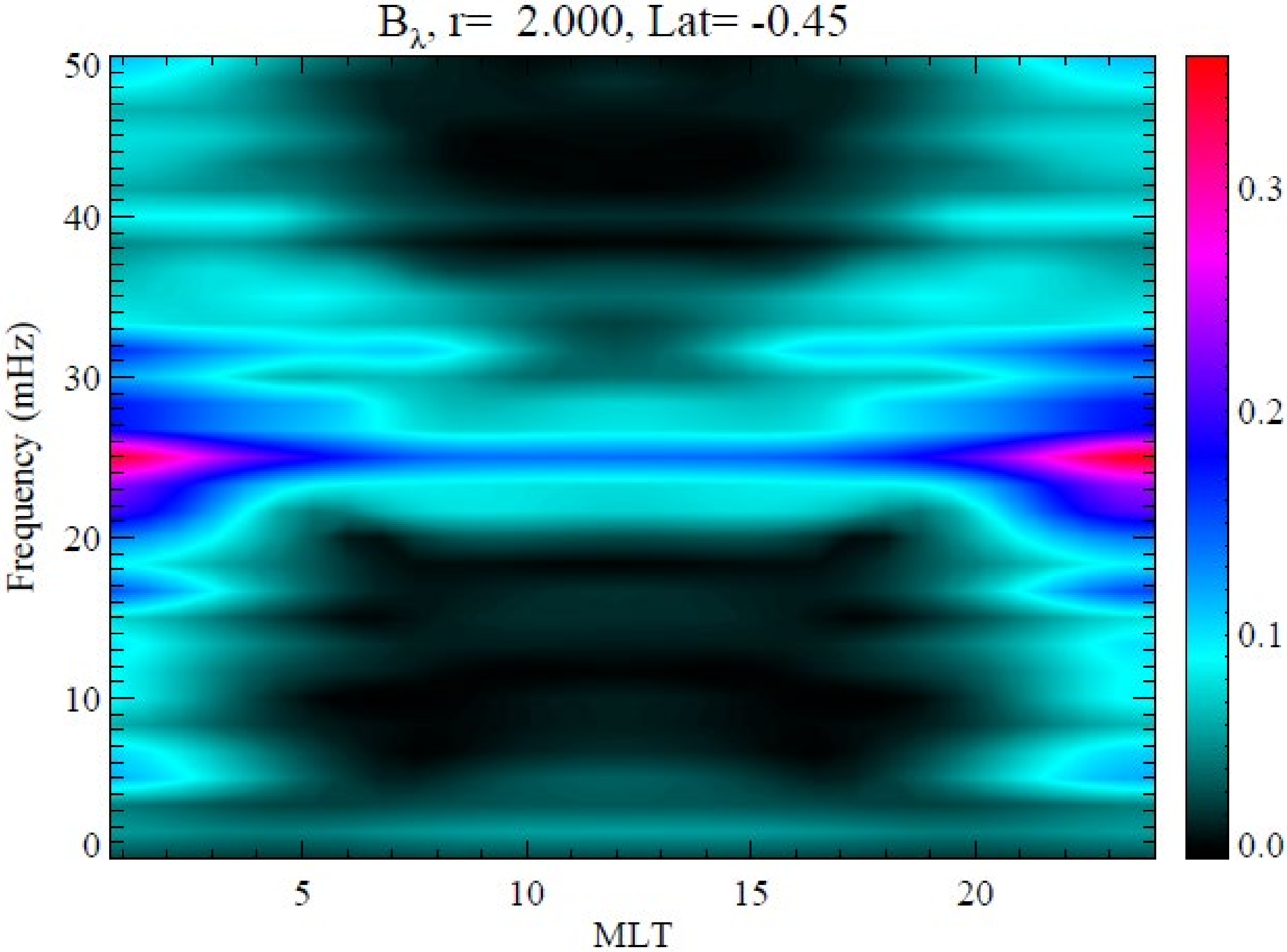

**Figure 10.** Azimuthal slice of the mode structure of the compressional $B_\lambda$ component. Driver was a collection of waves up to 50 mHz with random phases.